\documentclass[a4paper,superscriptaddress,twocolumn,pra,showkeys]{revtex4}
\usepackage{amsmath}
\usepackage{amssymb}
\usepackage{graphicx}
\usepackage{upgreek}
\usepackage{hyperref}
\usepackage[T1]{fontenc}

\begin{document}

\title{Fractal light from lasers}

\author{Hend Sroor}
\affiliation{School of Physics, University of the Witwatersrand, Private Bag 3, Wits 2050, South Africa}

\author{Darryl Naidoo}
\affiliation{School of Physics, University of the Witwatersrand, Private Bag 3, Wits 2050, South Africa}
\affiliation{CSIR National Laser Centre, P.O.\ Box 395, Pretoria 0001, South Africa}

\author{Steven W.\ Miller}
\author{John Nelson}
\affiliation{School of Physics and Astronomy, College of Science and Engineering, University of Glasgow, Glasgow G12 8QQ, UK}

\author{Johannes Courtial}
\email{Johannes.Courtial@glasgow.ac.uk}
\affiliation{School of Physics and Astronomy, College of Science and Engineering, University of Glasgow, Glasgow G12 8QQ, UK}

\author{Andrew Forbes}
\affiliation{School of Physics, University of the Witwatersrand, Private Bag 3, Wits 2050, South Africa}

\begin{abstract}
Fractals, complex shapes with structure at multiple scales, have long been observed in Nature: as symmetric fractals in plants and sea shells, and as statistical fractals in clouds, mountains and coastlines.
With their highly polished spherical mirrors, laser resonators are almost the precise opposite of Nature, and so it came as a surprise when, in 1998, transverse intensity cross-sections of the eigenmodes of unstable canonical resonators were predicted to be fractals [Karman \textit{et al.}, Nature \textbf{402}, 138 (1999)].
Experimental verification has so far remained elusive.  Here we observe a variety of fractal shapes in transverse intensity cross-sections through the lowest-loss eigenmodes of unstable canonical laser resonators, thereby demonstrating the controlled generation of fractal light inside a laser cavity.  We also advance the existing theory of fractal laser modes, first by 
predicting 
3D self-similar fractal structure around the centre of the magnified self-conjugate plane, second by showing, quantitatively, that intensity cross-sections are most self-similar in the magnified self-conjugate plane.
Our work offers a significant advance in the understanding of a fundamental symmetry of Nature as found in lasers.
\end{abstract}

\keywords{fractal laser modes, unstable laser, laser modes}

\maketitle

%% enable floats that take up most of a column/page, and stop them from piling up at the end of the document
%% see http://www.tex.ac.uk/FAQ-floats.html
\renewcommand{\topfraction}{.9}
\renewcommand{\bottomfraction}{.9}
\renewcommand{\textfraction}{.1}
\renewcommand{\floatpagefraction}{.9}
\renewcommand{\dbltopfraction}{.9}
\renewcommand{\dblfloatpagefraction}{.9}
\setcounter{topnumber}{9}
\setcounter{bottomnumber}{9}
\setcounter{totalnumber}{20}
\setcounter{dbltopnumber}{9}

%\significancestatement{Fractal geometry is the geometry of nature: the shape of trees and mountains is self-similar, and so a branch looks like a small tree and a rocky outcrop like a small mountain.
%With its highly polished spherical mirrors, a laser resonator is almost the precise opposite of nature, and so it came as a surprise when, in 1998, the intensity cross-sections of the light beams emitted from a class of laser resonator, namely unstable resonators, were predicted to be fractals.
%Here we provide the first experimental evidence for fractal light from canonical lasers and we extend the theory of these fractal laser eigenmodes, clarifying the self-similarity properties in two and three dimensions.
%}

%\usepackage{hyperref}
%\usepackage{bm}% bold math
%\usepackage{color}
%
%\newcommand{\rmi}{\mathrm{i}}
%\newcommand{\rmd}{\mathrm{d}}
%\newcommand{\bi}[1]{\mathbf{#1}}
\newcommand{\Note}[1]{\textcolor{red}{ #1}}

\section{Introduction}

The allure of fractals lies not only in their aesthetic beauty and the mathematical beauty of their self-similarity, but also in that such complexity can be achieved by very simple chaotic equations.
Nature seemingly utilises this as an engineering tool, with symmetric fractal structures appearing in many diverse forms, from romanesco broccoli to ammonite sutures and ferns, while statistical fractals are seen in salt flats, mountains, coastlines and clouds.
Popularised by Benoit Mandelbrot \cite{Mandelbrot-1982}, fractals can be thought of the the mathematical instance of ``plus ca change, plus c'est la meme chose'' (``the more things change, the more they stay the same'').

Light too can be fractal. The (dark) vortex lines in random light fields have fractal scaling properties \cite{OHolleran-et-al-2008}, and light's spatial (and spectral \cite{Lehman-Garavaglia-1999}) distribution can be directly made fractal by interaction with a fractal object, for example by emitting it from a fractal antenna \cite{Werner-Ganguly-2003}, by passing it through a fractal aperture \cite{Lehman-2001,Saavedra-et-al-2003}, or by resonating it in a cavity that contains a fractal scatterer \cite{Takeda-et-al-2004}. Perhaps more surprisingly, due to the fractal Talbot effect the light field behind a (non-fractal) Ronchi grating illuminated by a uniform plane wave evolves, on propagation, into a fractal~\cite{Berry-Klein-1996a,Berry-et-al-2001}.

A glance at intensity cross-sections through the eigenmodes of unstable canonical resonators (e.g.\ \cite{Siegman-1986-unstable-resonator-eigenmode-distributions}) reveals complex and fractal-looking structure, but the first suggestion that these eigenmodes are fractals came only in 1998 \cite{Karman-Woerdman-1998,Karman-et-al-1999}.
This is surprising, as canonical resonators are very simple, consisting of a pair of spherical mirrors and any apertures in the resonator.
Initially, the discussion of the mechanism involved mostly the round-trip magnification due to geometrical imaging by the spherical mirrors, which leads to similar patterns to appear at a cascade of different length scales, one of the hallmarks of fractals, but it also hinted at the role of diffraction, which gives rise to the ripples in the pattern in the first place \cite{New-et-al-2001}.
Clearly, without diffraction, successive magnifications would simply make any initial pattern increasingly uniform.
Detailed theoretical studies of these intensity distributions found them to be statistical fractals \cite{Karman-Woerdman-1998,Karman-et-al-1999,Berry-2001,Berry-et-al-2001,New-et-al-2001,Yates-New-2002,Loaiza-2005,Huang-et-al-2006}.
Upon magnification, statistical fractals look like the same \emph{type} of pattern, but not actually the same pattern.
Like in all physical fractals, the range of length scales over which this scaling behaviour holds (the scaling range) is limited \cite{Avnir-et-al-1998}, here by diffraction. 

\begin{figure}[ht]
	\begin{center} \includegraphics[width=1\linewidth]{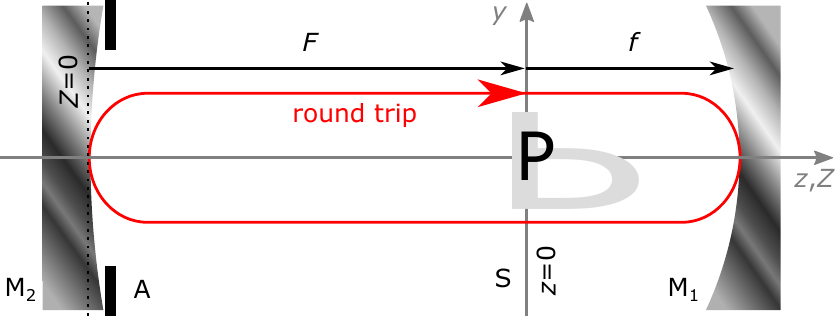} \end{center}
	\caption{Imaging inside an unstable canonical resonator.
	The two spherical mirrors, M$_1$ (focal length $f$) and M$_2$ (focal length $F$), perform geometric imaging.
	We define two longitudinal coordinates, $z$ and $Z$.
	Both the $z$ and the $Z$ axes coincide with the optical axis, but the $z=0$ plane coincides with the plane S, the magnified self-conjugate plane, and the $Z=0$ plane coincides with the plane of the mirror M$_2$. % (so $Z = z+F$).
	We use $z$ in our theoretical analysis, $Z$ in the experimental part.
	The transverse coordinates are $x$ (not shown) and $y$.
	Three-dimensional imaging during one round trip is indicated by an object in the shape of the letter ``P'' (shown in black) and its image, which looks like a horizontally elongated letter ``b'' (shown in grey):
		the ``P'' has turned into a ``b'' because the transverse magnification, $M$, is negative, and so the image of the ``P'' is upside-down;
		the ``b'' is horizontally elongated because the longitudinal magnification, $M_l$, is positive and its magnitude is greater than that of the transverse magnification.
		$A$ is an aperture immediately in front of M$_2$.
		The figure is drawn for a the particularly simple case of a confocal resonator (length $F + f$; the plane S then coincides with the common focal plane) with $M = -2$ and $M_l = +4$.}
	\label{resonator-figure}
\end{figure}

\begin{figure}[ht]% [H]
	\begin{center} \includegraphics[width=1\linewidth]{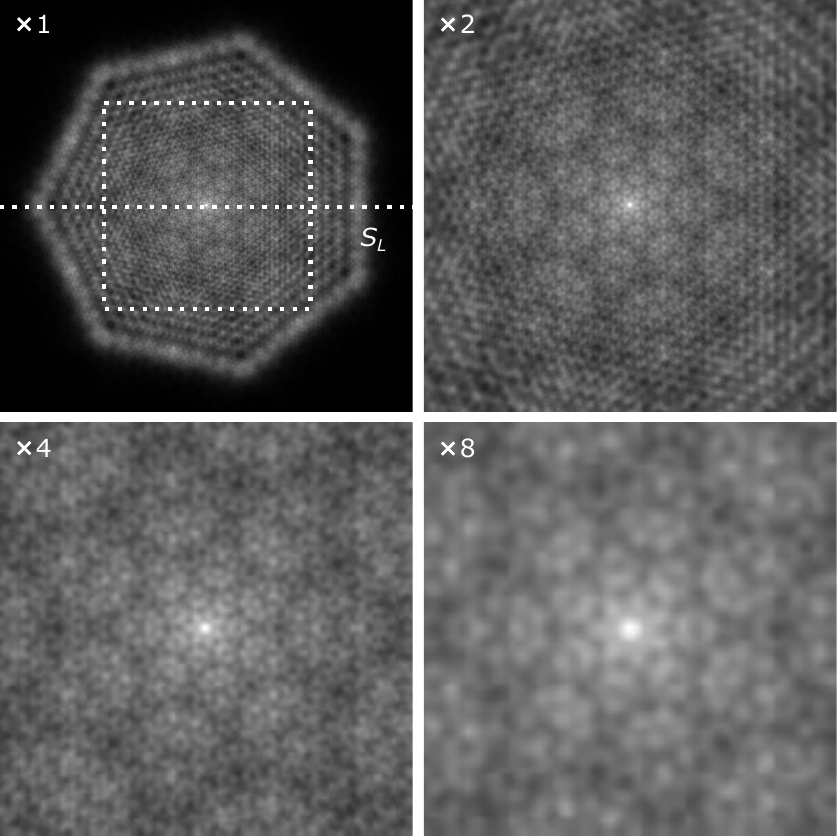} \end{center}
	\caption{\label{transverse-intensity-distribution-3D-resonator-figure}Self-similarity of the simulated lowest-loss eigenmode's intensity distribution in the magnified self-conjugate plane, S.
		The frames show the intensity after 20 round trips, starting with a uniform plane wave.
		The self-similarity of the pattern is demonstrated by showing its centre at different magnifications ($\times 2$, $\times 4$, $\times 8$), resulting in a similar pattern (rotated by $180^\circ$ after each magnification by a factor 2 due to the resonator's transverse magnification, $M$, being negative);
		the dotted white square in the centre of the frame marked $\times 1$ shows the outline of the area shown in the next frame.
		The horizontal dotted line is the orthographic projection of the lateral self-conjugate plane $S_L$, in which we demonstrate, below (see Fig.\ \ref{lateral-intensity-distribution-3D-resonator-figure}), the 3D self-similarity of the light.
		The figure is calculated for light of wavelength $\lambda = 632.8$\,nm in a resonator of the type shown in Fig.\ \ref{resonator-figure} with $F=16.5$\,cm, $f=8.25$\,cm, $M = -2$, and a seven-sided regular polygonal aperture of circumradius $r_0 = 2.4$\,mm.
		The beam's transverse cross-sections were represented by a $1024 \times 1024$ array of complex numbers sampled over a physical area of size 1\,cm$\times$1\,cm.
		For further details of the way this simulation, and indeed all the other simulations in this paper, was performed see App.~\ref{a:simulations}.
		%%Like all simulations in this paper, this simulation was performed using the open-source scalar-wave-optics simulator Young TIM~\cite{Leavey-Courtial-2016}.
	}
\end{figure}

Shortly after the original explanation for the fractal character of the eigenmodes of unstable resonators it was found that the role of diffraction was particularly simple in the plane that is geometrically imaged into itself with a magnification $M$ of modulus $|M| >1$, the \emph{magnified self-conjugate plane} \cite{Courtial-Padgett-2000b,Watterson-et-al-2003}.
In this plane, shown in Fig.\ \ref{resonator-figure}, the intensity distribution is a diffraction-limited self-similar fractal \cite{Courtial-Padgett-2000b}, with an example shown in Fig.\ \ref{transverse-intensity-distribution-3D-resonator-figure}.
The mechanism, called the \emph{monitor-outside-a-monitor effect} after a video-feedback analogy \cite{Courtial-et-al-2001,Leach-et-al-2003}, is that each round trip through the resonator, starting and finishing in the magnified self-conjugate plane, can be approximated as simple scaling by a factor $M$ of the initial beam and addition of the aperture diffraction pattern under spherical-wave illumination.  

After magnification, a part of a self-similar pattern looks not just to be the same pattern type as a corresponding, unscaled, part the pattern, but \emph{the same}.
Note that, with all physical fractals, this is only true over a finite range of sizes, here limited by the smallest size allowed by diffraction and the overall size of the beam.
Suitable choice of the resonator parameters has been predicted to lead to intensity distributions closely related to classic fractals such as the Weierstrass-Mandelbrot function, the Sierpinski gasket, and the Koch snowflake~\cite{Watterson-et-al-2003,Courtial-Padgett-2000b}.

Despite these early advances, experimental observations have been scarce. % experimental observation of fractals from lasers has remained elusive.
A pulse of light was injected into a passive canonical cavity and observed to evolve over a number of round trips into a fractal pattern \cite{Loaiza-et-al-2003}; curiously, that work was never published in a peer-reviewed journal.
Very recently, small areas containing fractal structure were found in the eigenmode of a non-canonical resonator comprising an array of microspheres sandwiched between planar mirrors \cite{Rivera-et-al-2018} --- the first observation of fractal structure in the eigenmode of an (active) laser.
These are the works that are most relevant to this study, but the relevance is limited as they either worked in a passive cavity and did not study the eigenmode in the magnified self-conjugate plane \cite{Loaiza-et-al-2003}, or in a different laser configuration altogether \cite{Rivera-et-al-2018}.
Further, % the prevalent diffraction argument has limited the scope of fractal dimensionality
the discussion was entirely limited to the light structure in transverse planes, resulting in what we will refer to as 2D fractals\footnote{Note that the dimension of the plane is, in general, different from the fractal dimension of such intensity distributions~\cite{Berry-et-al-2001}.}.

Here we experimentally verify the existence of self-similar fractal light from canonical lasers by observing the 2D intensity structure of laser light at the magnified self-conjugate plane inside the cavity and studying its self-similarity directly, rather than through the fractal dimension.
% We analyse the scaling factors (dimensionality of the fractals) and confirm the self-similar nature of the structure.
Further, we show that fractals can form in the three-dimensional (3D) intensity distributions of light in unstable canonical resonators.
We find that, around the centre of the magnified self-conjugate plane, this intensity distribution in 3D space is a self-similar fractal, albeit with different transverse and longitudinal characteristic scaling factors.
While we outline this structure in 3D space theoretically, the experimental verification remains an open task.

%with the role of diffraction outlined as an analogy between unstable resonators and the fractal Talbot effect \cite{Berry-Klein-1996a}: waves from different round trips analogous to waves from different unit cells~\cite{Berry-et-al-2001}.

%Successive round trips through unstable canonical resonators also result in fractal transverse intensity distributions;
%the lowest-loss eigenmodes of such resonators therefore have fractal intensity cross-sections \cite{Karman-Woerdman-1998,Karman-et-al-1999}.
%A number of authors studied the properties (fractal dimension etc.) of these intensity distributions and found them to be statistical fractals \cite{Karman-Woerdman-1998,Karman-et-al-1999,Berry-2001,Berry-et-al-2001,New-et-al-2001,Yates-New-2002,Loaiza-2005}.
%An analogy between unstable resonators and the fractal Talbot effect was also offered:
%the waves from different round trips in the former correspond to those from different unit cells in the latter~\cite{Berry-et-al-2001}.

%%%%%
%%%%%
\section{Theory}
%%%%%
%%%%%
\subsection{\label{s:transverseFractals}Transverse Fractals}
We start by reviewing the mechanism that leads to self-similar fractal structure at the self-conjugate plane in an unstable canonical resonator.
Without loss of generality, we restrict ourselves to confocal resonators, as these are particularly simple but at the same time representative of all canonical unstable resonators (with the same round-trip magnification, $M$, and the same Fresnel number \cite{Siegman-1986}).

Consider the example shown in Fig.~\ref{resonator-figure}.  In such a resonator, each mirror is spherical and so images like a lens, but in reflection.
During one round trip, i.e.\ reflection off both mirrors, the image produced by the first mirror is imaged again by the second mirror.
In stable resonators this imaging explains the eigenmodes' structural stability \cite{Forrester-et-al-2002}.
In unstable canonical resonators, one round trip images two ``self-conjugate'' planes back to their original positions, one with (transverse) magnification $M$ ($|M| \geq 1$), the other with magnification $1/M$ \cite{Courtial-Padgett-2000b}.
The former is the magnified self-conjugate plane, S, the latter the de-magnified self-conjugate plane, s.
In a confocal resonator, these planes are a focal distance on either side of the two mirrors (see Fig.\ \ref{resonator-figure}), and so the field in these planes forms a Fourier pair.
Geometrical imaging stretches, during every round trip through the resonator, the intensity distributions in the planes S and s by a factor $M$ and $1/M$, respectively.

Any apertures in the resonator simply add some diffractive ``decoration'' to this image.
After a number of round trips, the pattern is essentially unchanged between successive round trips (the complex amplitude cross-section is unchanged apart from complex factor representing a uniform phase change and loss), which means the field has settled into an eigenmode.

In our case, the lowest-loss eigenmode is reached after approx.\ 20 round trips. % (see Supplementary Information).
Once the eigenmode has formed, the decoration pattern is the same during successive round trips.
Once added, it gets magnified with the rest of the intensity distribution, which results in the presence of the decoration pattern in a number of sizes:
the pattern added during the most recent round trip is at the original size;
that added during the previous round trip is magnified by $M$;
that added two round trips ago is magnified by $M^2$;
and so on.
The presence of a pattern on such a cascade of length scales is a hallmark of self-similarity.
The mechanism outlined above is called the monitor-outside-a-monitor effect (MOM effect), named so because of analogies with video feedback~\cite{Courtial-et-al-2001,Leach-et-al-2003}.

% We also measured the fractal dimension of several intensity cross-sections, using the standard Fourier-space technique \cite{Yates-New-2002}.
% This is shown in Fig.\ \Note{Do this analysis!  I will play a bit in Mathematica, and try to create a graph like Fig.\ 8.}
% Fractal dimension is more complicated \cite{Berry-et-al-2001}.

%\hend{A simple verification for such mechanism has been demonstrated by simply simulate the resonator represented in Fig.~\ref{resonator-figure} with $F= 16.5$ cm, $f=8.25$ cm and $M=-2$. We started with a uniform plane wave of wavelength 632.8 nm. A seven-sided polygen aperture of circumradius 2.4 mm was placed before M$_2$.    
%Fig.\ \ref{transverse-intensity-distribution-3D-resonator-figure} illustrates the eigenmode's intensity cross-section profile in the plane~S. The the self-similarity of the eigenmode was verified by detecting the starting decoration pattern at different scales by zooming into the centre of the profile cross-section by factor $M^i$ where $i={0,1,2,3}$ }

%%%%%
%%%%%
\subsection{3D Fractals}
\begin{figure}[ht]
	\begin{center} \includegraphics[width=1\linewidth]{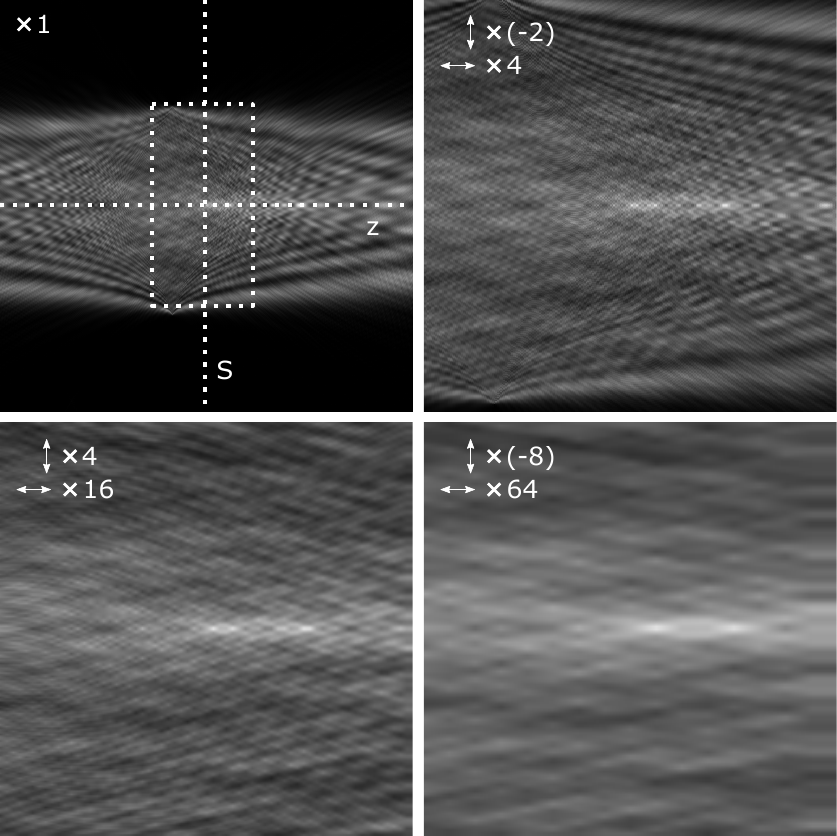} \end{center}
	\caption{Self-similarity of the intensity distribution in a lateral self-conjugate plane, $S_L$, which contains the optical axis and intersects the plane S horizontally in Fig.~\ref{transverse-intensity-distribution-3D-resonator-figure}.
		The vertical dotted line is the orthographic projection of the plane S.
		Vertically, the plots are centred on the optical axis, $z$.
		The beam is the same as that shown in Fig.\ \ref{transverse-intensity-distribution-3D-resonator-figure}.
		After each magnification horizontally by a factor 4 and vertically by $-2$, the pattern looks similar again, which is shown for different magnifications.
		The dotted box in the centre of the frame marked $\times 1$ marks the outline of the area shown in the next frame ($\times (-2)$ vertically, $\times 4$ horizontally).
		The $\times 1$ frame represents a physical area of size 2\,m (horizontally) by 10\,mm (vertically).}
	\label{lateral-intensity-distribution-3D-resonator-figure}
\end{figure}

%%%%%
%%%%%
For the same resonator, Fig.~\ref{lateral-intensity-distribution-3D-resonator-figure} shows a lateral intensity distribution around the centre of the self-conjugate plane S.
This lateral intensity distribution shows some signs of self-similarity:
if the pattern is stretched by $M$ in the direction representing the transverse direction, and by a factor $M^2$ in the longitudinal direction, the pattern's centre (which is the point where the plane S intersects the resonator's optical axis) looks similar to what it was before magnification.

\begin{figure}[ht]
	\begin{center} \includegraphics{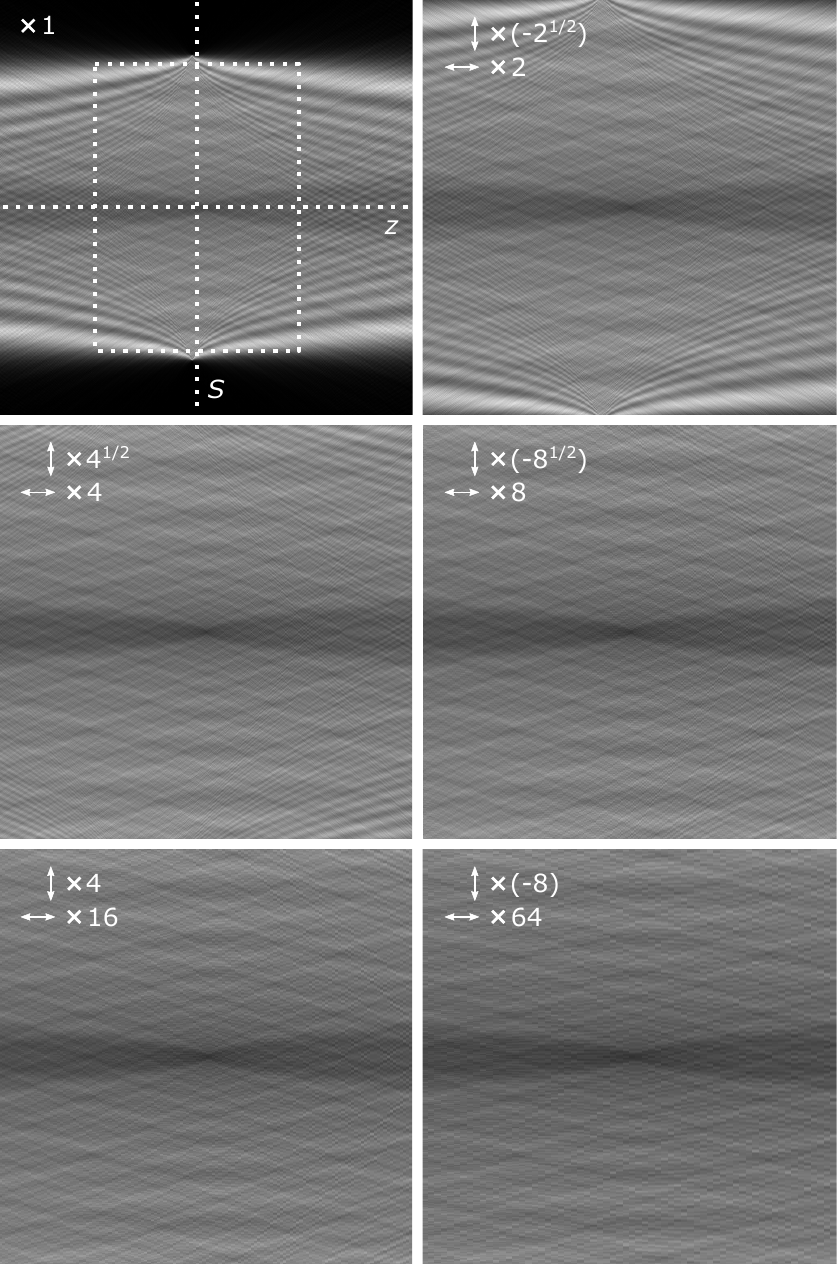} \end{center}
	\caption{Self-similarity of the intensity distribution in the lateral plane of a strip resonator of the type shown in Fig.\ \ref{resonator-figure}.
		The different frames show the centre of the intensity distribution, successively magnified by a factor $M$ in the vertical direction and by $M^{2}$ in the horizontal direction.
		The $\times 1$ frame represents a physical area of size 20\,m (horizontally) by 2.82\,cm (vertically), centred on the magnified self-conjugate plane and the optical axis in the horizontal and vertical direction, respectively.
		The dotted box shown in the top left frame outlines the area shown in the top right frame.
		The figure was calculated for light of wavelength $\lambda = 632.8$\,nm, resonator parameters $F = 70.7$\,cm and $f = 50$\,cm (corresponding to transverse magnification $M = -\sqrt{2}$), and the aperture $A$ was a slit of width 2.08\,cm.
		Each beam cross-section was represented on 4096-element array of complex numbers, representing a physical width 4\,cm.
	}
\label{lateral-intensity-distribution-2D-resonator-figure}
\end{figure}

\begin{figure}[ht]
	\begin{center} \includegraphics[width=\linewidth]{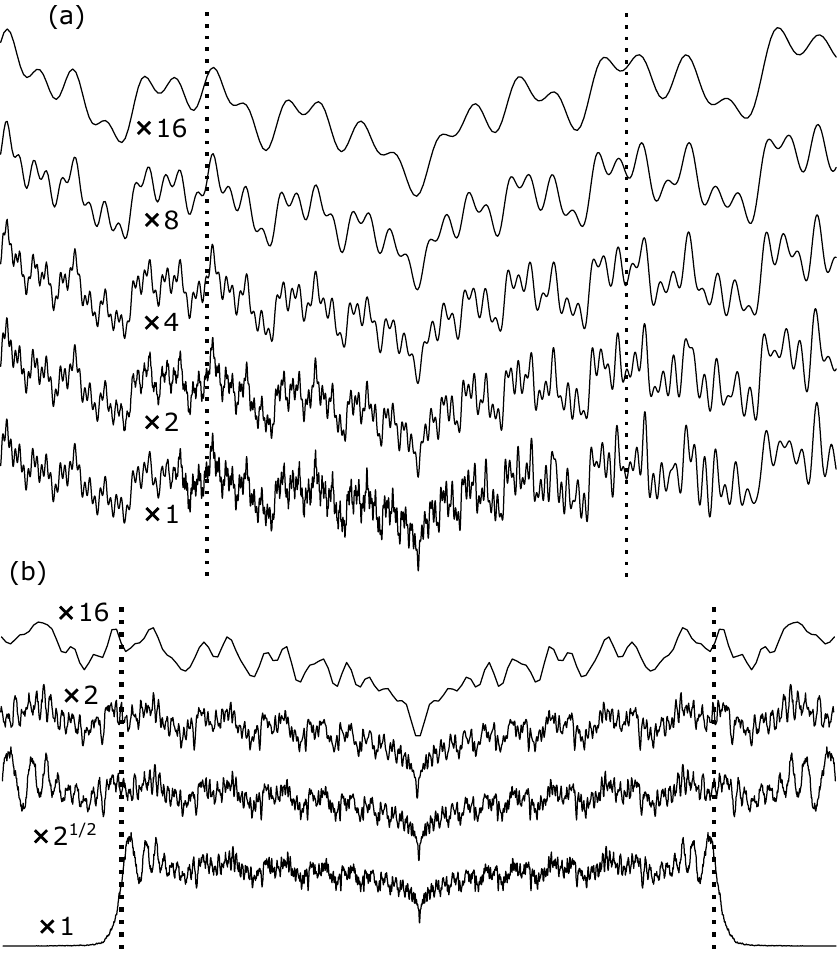} \end{center}
	\caption{Axial~(a) and transverse~(b) intensity cross-section through the field around the self-conjugate point at the centre of the plane S in the strip resonator from Fig.\ \ref{lateral-intensity-distribution-2D-resonator-figure}.
		Like in Figs \ref{transverse-intensity-distribution-3D-resonator-figure}, \ref{lateral-intensity-distribution-3D-resonator-figure} and \ref{lateral-intensity-distribution-2D-resonator-figure}, the self-similarity is demonstrated by successive magnifications, each of which stretches the part of the curve between the vertical dotted lines to the full width.
		The width of the curves marked $\times 1$ represents a physical length of 2\,m (a) and 2.08\,cm (b).
		The intensity range represented by the different curves has been adjusted so that corresponding features in the curves have roughly the same vertical size.}
		\label{line-intensity-distributions-2D-resonator-figure}
\end{figure}

This self-similarity can be seen much clearer in Fig.~\ref{lateral-intensity-distribution-2D-resonator-figure}, which was calculated for a strip resonator, i.e.\ a resonator that is invariant in one transverse direction.
It can therefore be treated as a 2D resonator with only one transverse direction, which means that, along that transverse direction, the light field can be represented in computer simulations by a much greater number of grid points without increasing memory or complexity requirements.
% For example, if memory allows $n$ data points to represent a transverse light-field cross-section, 
This in turn allows an increase in the Fresnel number by increasing the aperture size, resulting in a lateral intensity cross-section with significantly more detail.

For that same strip resonator, Fig.~\ref{line-intensity-distributions-2D-resonator-figure} compares the intensity cross-sections along the transverse direction in the plane S with that along the resonator's optical axis.
The intensity cross-section along the optical axis is not symmetrical with respect to the position of the plane S, whereas that in the plane S is symmetric with respect to the position of the optical axis.
Irrespective of this complication, both curves are strikingly self-similar. % the self-similarity properties of these two curves are strikingly similar.

This observation can be explained as follows.
Spherical mirrors (and lenses) image not only any transverse plane into a corresponding transverse plane, they image any point into a corresponding point.
For light initially travelling to the right in the resonator shown in Fig.~\ref{resonator-figure}, any lateral plane that includes the optical axis is being imaged into itself, as is the magnified self-conjugate plane S; no other planes are being imaged into themselves (but other surfaces are, specifically the paraboloids $z = a r^2$, where $z$ and $r$ are cylindrical coordinates as shown in Fig.\ \ref{resonator-figure} and $a$ is an arbitrary constant).
One point is imaged into itself (``self-conjugate point''), namely the intersection of the self-conjugate plane S with the optical axis.
The volume around this point is also imaged into itself, but the image is distorted as the longitudinal and transverse magnifications are different (the longitudinal magnification is the square of the transverse magnification) and both change with position.
(Similar statements are true for light initially traveling to the left, but we do not consider these here.)
Close to the self-conjugate point, the longitudinal magnification is constant.
This imaging of the volume around the centre of the plane S is indicated in Fig.~\ref{resonator-figure}.

As before, the effect of any apertures in the system is the addition of a diffractive decoration pattern, which is now 3D.
In a 3D extension of the MOM effect, this pattern gets added to the field during each round trip and magnified during each subsequent round trip, again resulting in its presence on a cascade of length scales, complicated and enriched by the different characteristic stretch factors in the longitudinal and transverse directions.

%%%%%
%%%%%
\subsection{Self-similarity of transverse fractals}
%%%%%
%%%%%
The mechanism for the emergence of fractals in the transverse intensity cross-sections, described in section \ref{s:transverseFractals}, suggests that the cross section is most self-similar in the magnified self-conjugate plane, but this has never been tested quantitatively.

Here we provide the first quantitative evidence for this argument.
For the intensity cross-section in one transverse plane at a time we calculate the normalised squared Euclidean distance, $d^2$, between the centre of this intensity cross-section and the centre of the same intensity cross-section, stretched by a factor $M$ (see App.\ \ref{a:nsed} for details).
This is a measure of the difference between the stretched and unstretched centre of the intensity cross-section, and $-d^2$ is therefore a measure of their similarity.
We then plot $-d^2$ as a function of the $z$ coordinate of the transverse plane, defined as the Cartesian coordinate aligned with the optical axis such that the plane $z=0$ is the magnified self-conjugate plane (Fig.\ \ref{resonator-figure}). %  behind the magnified self-conjugate plane.
From the above argument we expect a peak at $z = 0$, that is, in the magnified self-conjugate plane~S.

\begin{figure}[ht]
\begin{center} \includegraphics{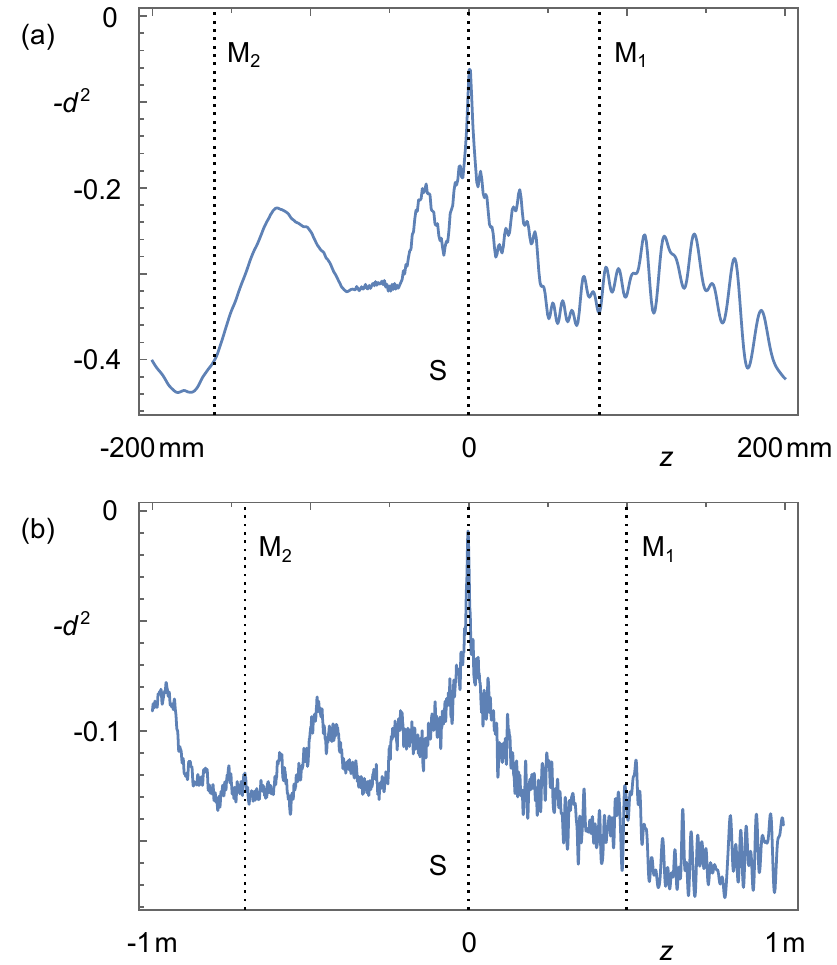} \end{center}
\caption{\label{f:selfSimilarity}Evolution of the self-similarity of transverse intensity cross-sections upon propagation for the eigenmodes shown in Figs \ref{transverse-intensity-distribution-3D-resonator-figure} and \ref{lateral-intensity-distribution-3D-resonator-figure} (a) and in Figs \ref{lateral-intensity-distribution-2D-resonator-figure} and \ref{line-intensity-distributions-2D-resonator-figure}~(b).
The self-similarity is quantified here by $-d^2$, the negative of the normalised squared Euclidean distance between the central rectangle of the intensity cross-section in the transverse plane with the given $z$ coordinate, and the central rectangle of the same size of the intensity cross-section after being stretched in the transverse directions by a factor $M$.
The magnification was $M = -3$ in (a) and $M=-2$ in~(b).
The width and height of the central rectangle were arbitrarily chosen to be $1/4$ of the width and height of the calculated intensity cross-section.
The dotted vertical lines indicate the planes of the mirrors, M$_1$ and M$_2$, and the magnified self-conjugate plane, S.}
\end{figure}

Fig.\ \ref{f:selfSimilarity} shows such curves, calculated for the two eigenmodes discussed earlier, namely that of a resonator with a heptagonal aperture and transverse magnification $M=-3$ (Figs \ref{transverse-intensity-distribution-3D-resonator-figure} and \ref{lateral-intensity-distribution-3D-resonator-figure}), and that of a strip resonator with transverse magnification $M=-\sqrt{2}$ (Figs \ref{lateral-intensity-distribution-2D-resonator-figure} and \ref{line-intensity-distributions-2D-resonator-figure}).
The expected peak at $z = 0$ is clearly visible and, especially in the case of the strip resonator, reaches close to $d^2 = 0$, proving the near-exact --- but diffraction-limited --- self-similarity of the intensity cross-section in the magnified self-conjugate plane.

Note that these results can be replicated with other measures of difference between images, and we did this with Euclidean distance and image Euclidean distance \cite{Wang-et-al-2005a} (IMED, which we calculated only for the strip-resonator eigenmode).
These gave a curve with a different shape to those shown in Fig.\ \ref{f:selfSimilarity}, but always produced a very prominent peak in the magnified self-conjugate plane.
% \Note{This is an important point that has previously been overlooked, where non-lasing cavities in a time-gated ring-down configuration were seeded by specific patterns and the output plane measured directly without consideration for the magnified self-conjugate plane \cite{Loaiza-2005}.}

\begin{figure}[ht]
	\begin{center} \includegraphics[width=\columnwidth]{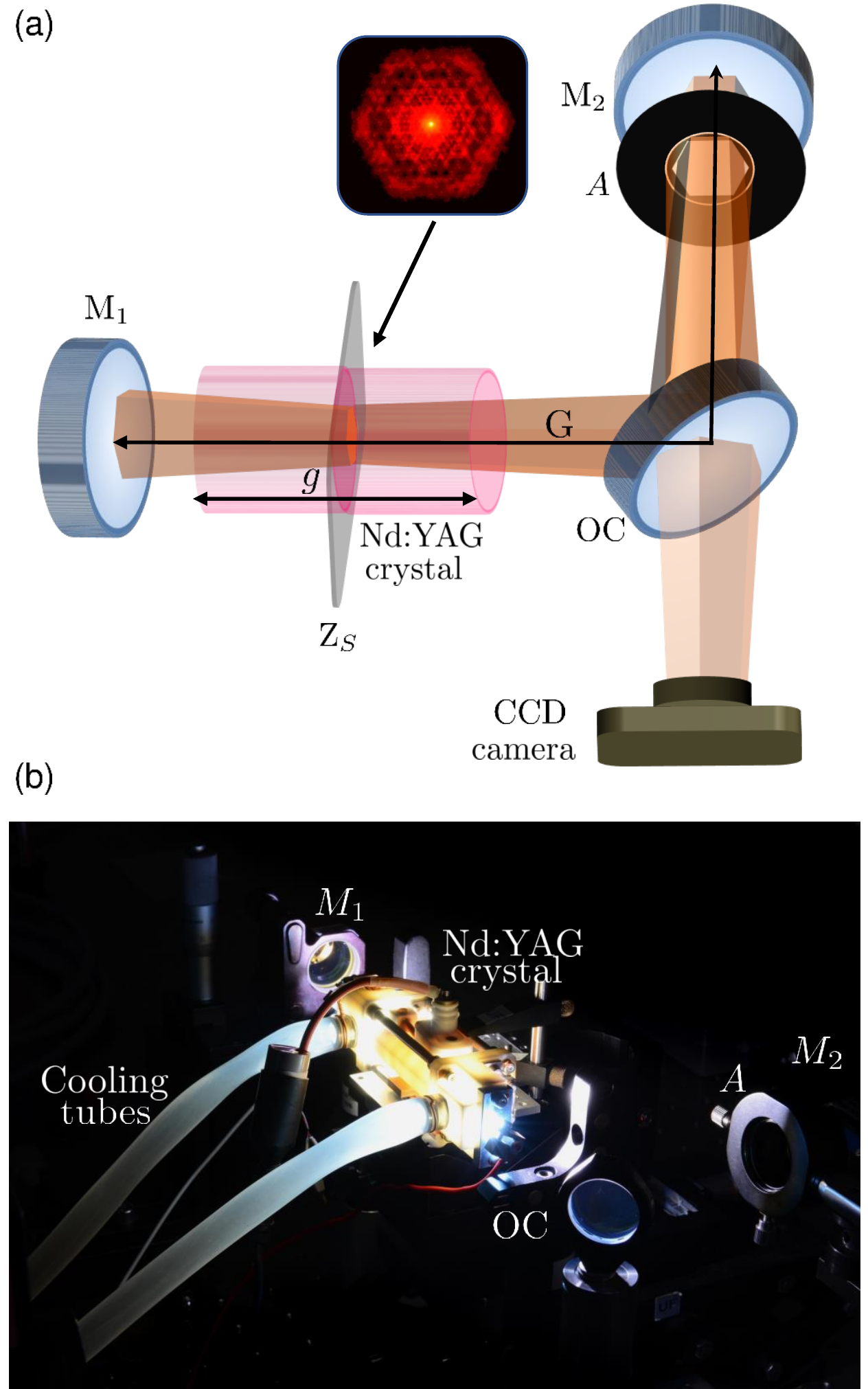} \end{center}
	\caption{Experimental setup, (a)~schematic and (b)~photo with the laser running.
	(a)~The cavity comprised two concave mirrors, M$_1$ (radius of curvature $R_1$, corresponding to focal length $f = R_1/2$) and M$_2$ (radius of curvature $R_2$, focal length $F = R_2/2$), and an output coupler (OC) angled at 45$^\circ$ with a 99.8\% reflectivity.
	The geometrical length of the cavity was $G$, %chosen to be $R_1/2 + R_2/2 = f + F$;
	that of the Nd:YAG gain medium was $g$. % , was positioned at the centre of the magnified self-conjugate plane S.
	S is the magnified self-conjugate plane.
	A polygonal aperture (A) was positioned in front of M$_2$.
	The intensity cross-section in the self-conjugate plane S was captured outside the cavity at a distance $R_2/2$ from M$_2$ using a CCD camera.
	(b)~Clearly visible is the gain-medium assembly (bright yellow; centre) and the mounts for the end mirrors and the output coupler.
		The two tubes (light blue; bottom left) connecting to the gain-medium assembly are cooling tubes for the gain medium.
		}
	\label{setup}
\end{figure}

%%%%%%%%%%%%%%%%%%%%%%%%%%%%%%%%%
\section{Experiment}
\label{s:experiment}
%%%%%%%%%%%%%%%%%%%%%%%%%%%%%%%%%
We constructed a laser, sketched in Fig.~\ref{setup}, consisting of a flash-lamp pumped Nd:YAG gain medium (6.35\:mm $\times$ 76\:mm) inside an L-shaped, confocal, unstable cavity comprising two concave, high-reflectivity, spherical end mirrors, M$_1$ and M$_2$, and a 45$^\circ$ output coupler positioned at the apex of the L.
The radius of curvature of mirror M$_1$ is $R_1$, that of M$_2$ is $R_2$; their focal lengths are respectively $f = R_1/2$ and $F = R_2/2$.
The geometrical length of the resonator is $G$.
% As the cavity is confocal (the cavity was set up to have length $f + F$), the plane S coincides with the common focal plane.
% In order to minimise unwanted aperture effects, the gain medium was centred on S.
An aperture in the shape of a regular hexagon was positioned in front of the end mirror with the greater focal length, M$_2$.

Like that sketched in Fig.\ \ref{resonator-figure}, our cavity is canonical and unstable, and as such it contains magnified and de-magnified self-conjugate planes.
Unlike that sketched in Fig.\ref{resonator-figure}, our cavity was not confocal, and so the positions of the self-conjugate planes did not simply coincide with the focal planes, but their $Z$ coordinate was calculated from the geometric-imaging properties of the cavity (see App.\ \ref{a:S-plane}).
% During each round trip in the resonator, light was therefore imaged geometrically by the resonator's two curved mirrors and the aperture added diffractive decoration.
The output beam was captured using a CCD camera (Spiricon SPU260 BeamGage), placed in an image plane of S.
Note that the field depends on the propagation direction (even in stable canonical resonators \cite{Naidoo-et-al-2012}), and the camera has been placed to record the image in the magnified self-conjugate plane, S, not the de-magnified self-conjugate plane, s, which corresponds to the same plane but the opposite propagation direction.

One of the intensity cross-sections observed in the magnified self-conjugate plane is shown as part of Fig.\ \ref{setup}(a).
It has the expected complex structure characteristic of a fractal.
In almost all images we note an unexpectedly bright central intensity peak.
We speculate that it is due to a low divergence mode that is also able to lase in the cavity.

\begin{figure*}[ht]
	\begin{center} \includegraphics[width=\linewidth]{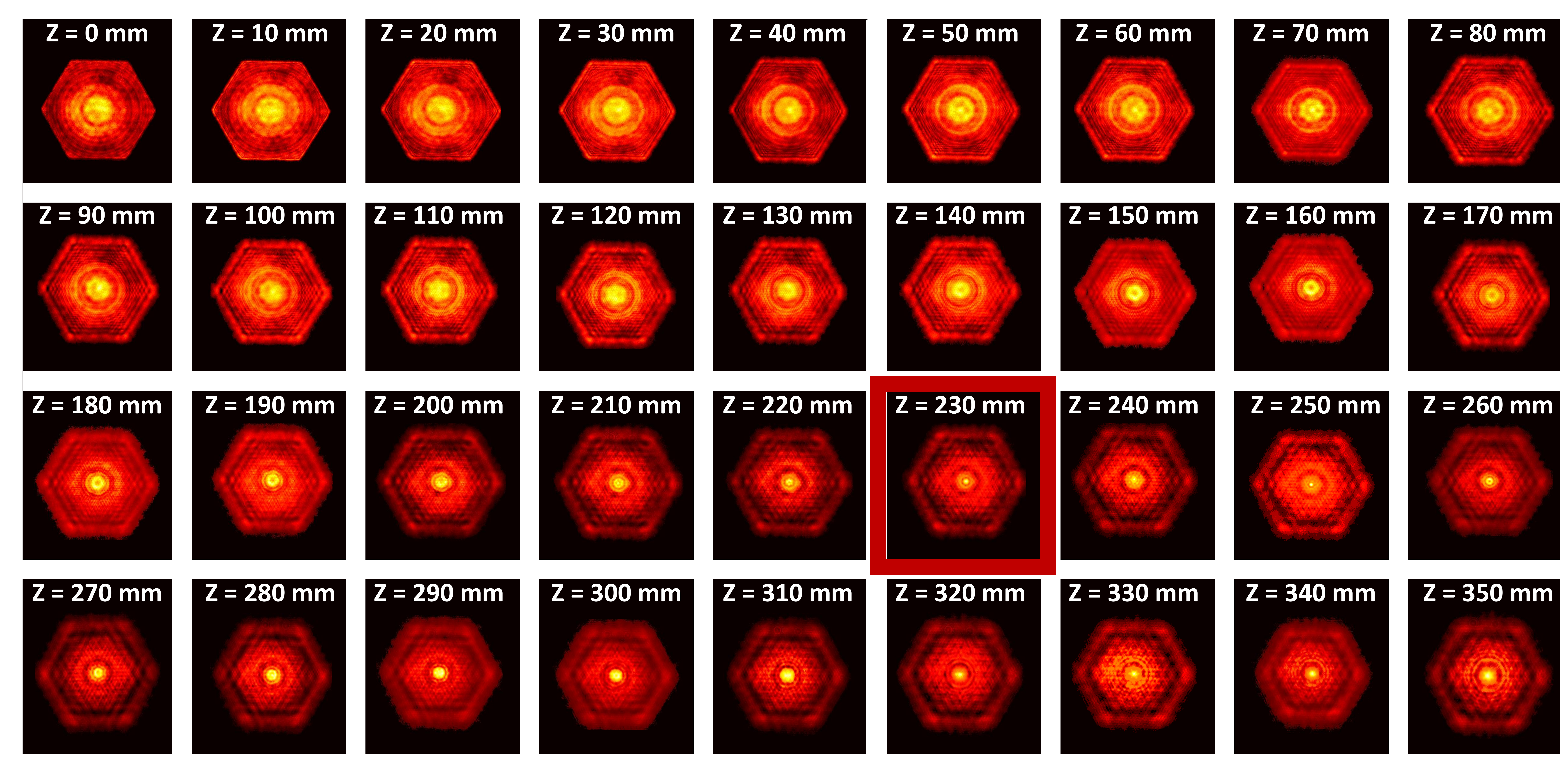} \end{center}
	\caption{Evolution of the eigenmode intensity cross-section between mirrors M$_2$ ($Z = 0\,\mathrm{mm}$) and M$_1$ ($Z = 343\,\mathrm{mm}$).
	The magnification was $M$ = 3.32 ($F=250$ \,mm, $f = 75$\,mm) with an aperture of circumradius $r_0 = 2\,$mm.
	The geometrical length of the cavity was $G = 343$\,mm, corresponding to an effective length of $L = 308.5$\,mm due to the refractive index of the gain medium (see App.~\ref{a:S-plane}).}
	\label{fracz}
\end{figure*}

\begin{figure}[ht]
\begin{center} \includegraphics[width=\linewidth]{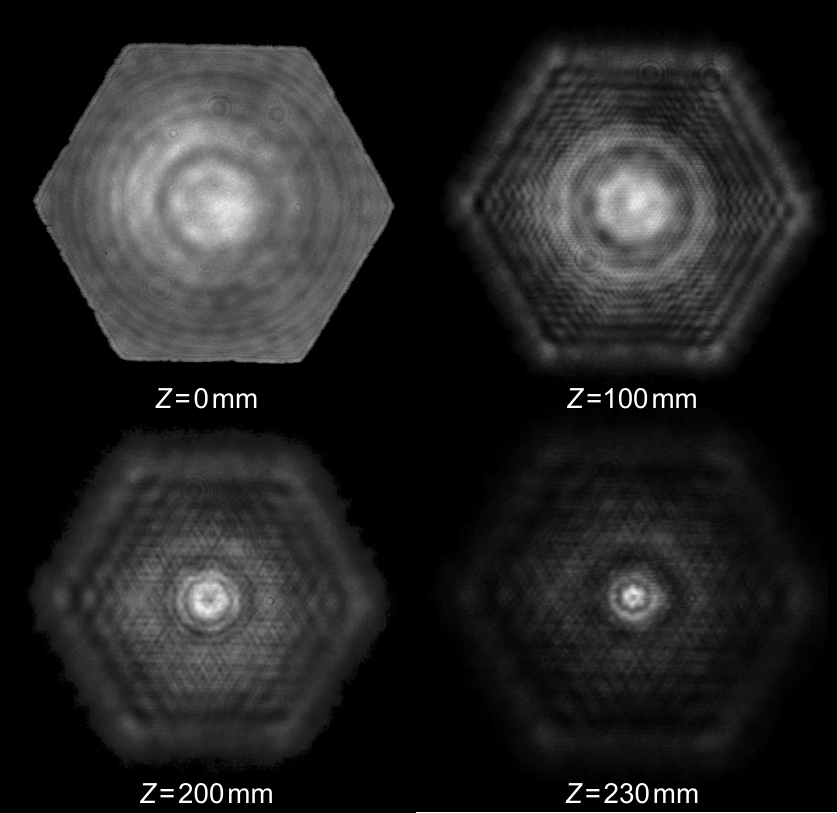} \end{center}
\caption{A few of the transverse intensity patterns recorded inside a resonator for parameters $F = 250\,\mathrm{mm}$, $f = 75\,\mathrm{mm}$, and $G = 343\,\mathrm{mm}$.
The self-conjugate plane S is positioned at $Z = Z_S = 231.8\,\mathrm{mm}$. %, including close to the self-conjugate plane. 
}
\label{frac3}
\end{figure}

For one set of parameters, namely $F = 250\,\mathrm{mm}$, $f = 75\,\mathrm{mm}$, $G = 343\,\mathrm{mm}$, and a hexagonal aperture A of circumradius $r_0 = 2.5\,\mathrm{mm}$ we recorded intensity cross-sections in a number of transverse planes across the resonator (Fig.\ \ref{fracz}), % (see Sec.\ \ref{a:evolution}), 
with examples shown in Fig.~\ref{frac3}.
The magnified self-conjugate plane is located at $Z_S = 231.8\,\mathrm{mm}$ (calculated using Eqn [\ref{e:Z_S}]), the round-trip transverse magnification is $M = -3.32$~(calculated from Eqn~[\ref{e:M}]).

\begin{figure*}[ht]
\begin{center} \includegraphics{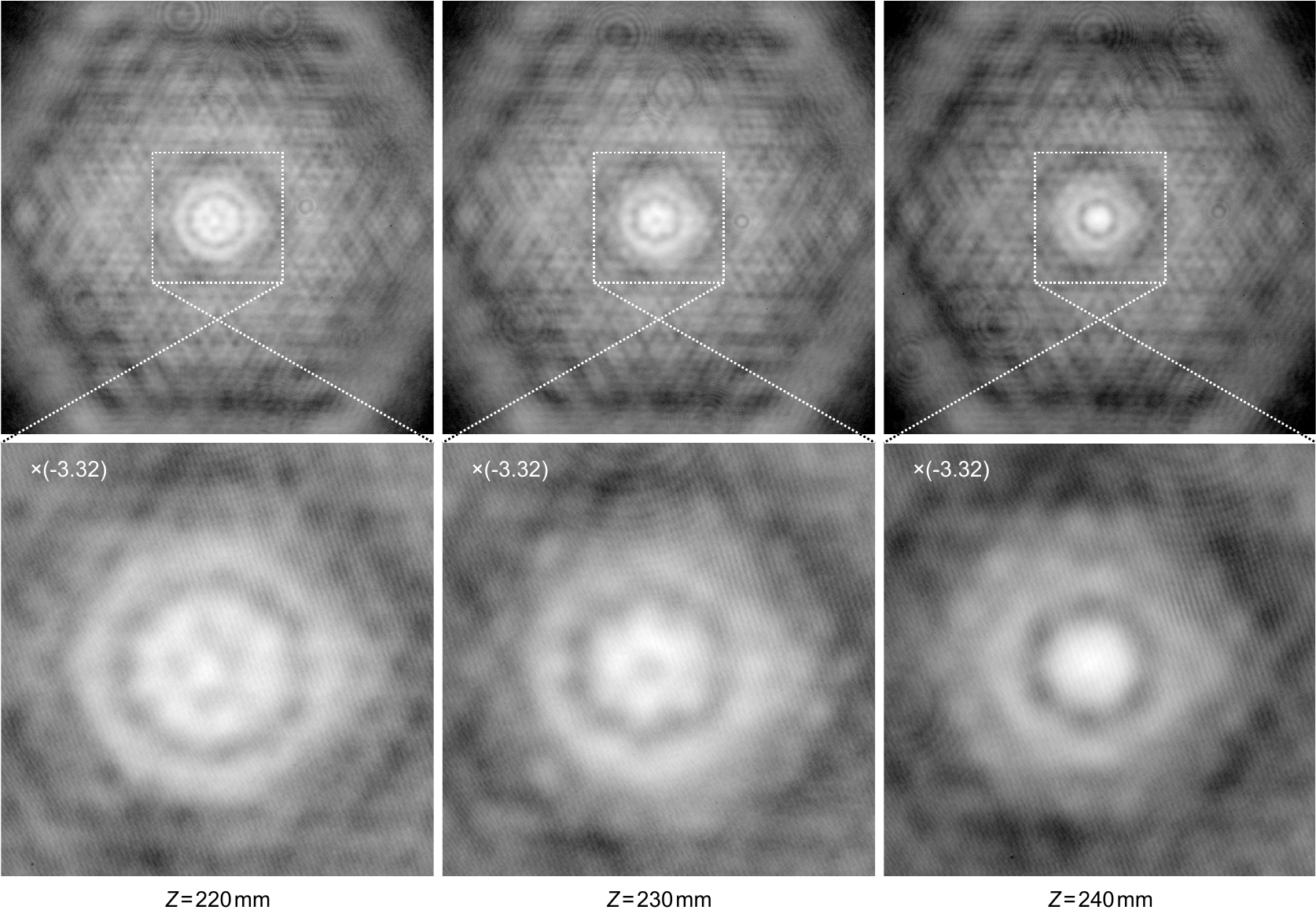} \end{center}
\caption{\label{f:selfSimilarityZ23To26}Self-similarity of the intensity cross-sections in different transverse planes near the predicted position of the self-conjugate plane, $Z = Z_S = 231.8\,\mathrm{mm}$.
The images in the top row show the central $\approx 4\,\mathrm{mm} \times 4\,\mathrm{mm}$ of the experimentally recorded intensity cross-sections in the transverse planes $Z = 220\,\mathrm{mm}$~(left), $Z = 230\,\mathrm{mm}$~(center), and $Z = 240\,\mathrm{mm}$~(right).
The images in the bottom row show the centers of the corresponding top-row images, stretched by a factor $M = -3.32$, that is, stretched by a factor $3.32$ and rotated by $180^\circ$.
To show structure over a wider intensity range, the brightness of each pixel is proportional to the logarithm of the recorded intensity.}
\end{figure*}

Fig.\ \ref{f:selfSimilarityZ23To26} allows visual assessment of the self-similarity of the intensity cross-sections recorded in three transverse planes near the self-conjugate plane S.
When stretched by the transverse magnification $M$, the centers of the intensity cross-sections show some similarity to the unstretched intensity cross-sections; for example, dark, centered, hexagons or circles of approximately the same size are present in the stretched and unstretched intensity cross-sections, especially in the planes $Z = 230\,\mathrm{mm}$ and $Z = 240\,\mathrm{mm}$, which are closest to the self-conjugate plane~S.

\begin{figure}[ht]
\begin{center} \includegraphics{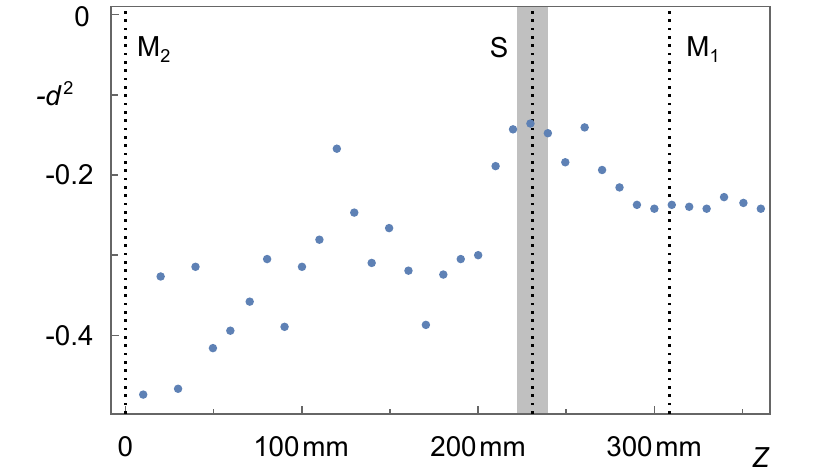} \end{center}
\caption{\label{f:selfSimilarityExperiment}Evolution of the self-similarity upon propagation of measured transverse intensity cross-sections.
Like in Fig.\ \ref{f:selfSimilarity}, the plot shows the negative normalised squared Euclidean distance, $-d^2$, as a function of axial coordinate $Z$.
The predicted position of the self-conjugate plane S, $Z = Z_S = 231.8\,\mathrm{mm}$, is indicated by a dotted black line surrounded by a grey area representing experimental uncertainties. % caused by thermal effects, small errors on mirror curvatures and distances. % behind the predicted position of the S plane ($z = 250$ mm).
% \AF{theoretical S plane of $z = 250$ mm.}
% \AF{The planes of the mirrors, M$_1$ and M$_2$, are indicated by vertical dotted lines while the estimated position of the S plane is shown as a grey band.}
The planes of the mirrors, M$_1$ and M$_2$, and the self-conjugate plane, S, are indicated by vertical dotted lines.
%\Note{I am not quite sure about this.  I think we need to justify this experimental uncertainty better, or remove it.  We do not currently use it in the main text.}
%Experimental equivalent of Fig.\ \ref{f:selfSimilarity}.
%There is no equivalent, sharp, trough, just a general depression around the expected self-conjugate plane, marked as $z = 0$.
%(b)~Dependence of the self-similarity as a function of magnification, $M$, for different $z$ planes in the depression shown in (a).
%For the ``correct'' self-conjugate plane, we expect this curve to have a minimum at $M=-3.33$.
%It can be seen that none of them does, but the planes with $z < 0$ have their dip at $M>-3.33$, that with $z > 0$ at $M<-3.33$.
%From this perspective, the plane $z=0$ should be about right, but it doesn't have a dip at all around $M=-3.33$.
}
\end{figure}

% We recorded intensity cross-sections in a number of transverse planes across the resonator, shown in Fig.\ \ref{fracz} in Sec.\ \ref{a:evolution}.
We analysed the recorded intensity cross-sections quantitatively by evaluating their self-similarity and plotting the evolution of this self-similarity upon propagation --- the experimental analog to the curves shown in Fig.\ \ref{f:selfSimilarity}.
% As in the plot calculated from simulated data (Fig.\ \ref{f:selfSimilarity}), the mirror planes are in no way distinguished in the plot of the self-similarity
The result is shown in Fig.\ \ref{f:selfSimilarityExperiment}.
The mirror planes are in no way special, which was also the case in the plot calculated from simulated data (Fig.\ \ref{f:selfSimilarity}).
The self-similarity is greatest around the expected position of the plane S, but the sharp peak visible in the curves shown in Fig.\ \ref{f:selfSimilarity} is absent.
One possible explanation for the absence of this peak is that, due to the theoretical sharpness of the peak (Fig.\ \ref{f:selfSimilarity}), we did not record the intensity in a plane close enough to S despite sampling a plane only $1.8\,\mathrm{mm}$ from S.
Another possible explanation is that the lack of the sharp peak is caused by experimental imperfections.
One type of such experimental imperfections, including thermal effects and errors on mirror curvatures and distances, could have resulted in S being located further than expected from the nearest plane that was sampled.
Another type, experimental imaging imperfections, most likely due to the effect of the gain medium (which images only in the paraxial limit), could have led to fine detail in the intensity cross-section missing.

\begin{figure*}[ht]
	\begin{center} \includegraphics[width=\linewidth]{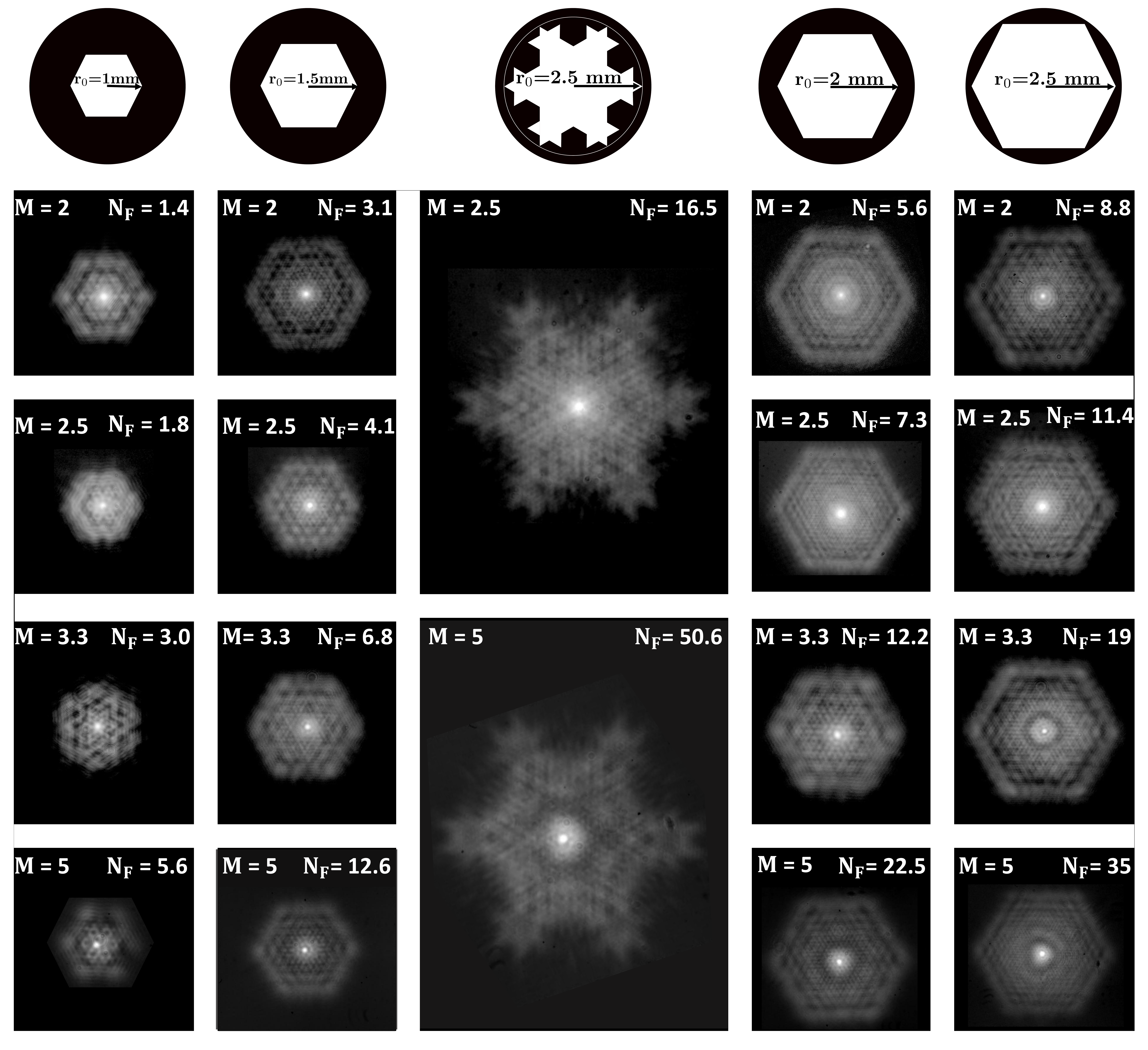} \end{center}
	\caption{Collage of experimentally obtained intensity cross-sections measured close to the self-conjugate plane S inside a variety of laser cavities.
	The edge of the aperture in the central column is that of the 3rd iteration of the Koch snowflake (the first iteration being an equilateral triangle), the other apertures are regular hexagons.
	$r_0$ is the circumradius of the apertures, $M$ is the approximate transverse round-trip magnification, and $N_F$ is the Fresnel number.
	Like in Fig.\ \ref{f:selfSimilarityZ23To26}, brightness is proportional to the logarithm of intensity.
	}
	\label{collage}
\end{figure*}

%\begin{figure}[ht]
%\begin{center} \includegraphics[width=4cm]{KochAperture.jpg} \end{center}
%\caption{Photo of the Koch-snowflake aperture used in Fig.\ \ref{collage}.
%The edge of the aperture is in the shape of the 3rd iteration of the Koch snowflake (the first iteration being an equilateral triangle).}
%\label{f:KochAperture}
%\end{figure}

%The evolution of the fractal structure along the length of the laser resonator reveals \Note{xxxx say what we find xxxx,} as shown in Fig.~\ref{} \Note{[xxxxx(figure for different z planes with their fractal dimensions)]}.  We find no fractal structure at the position of the apertured mirror, confirming the erroneous null result found by others \cite{Loaiza-et-al-2003,Loaiza-2005}.
 
%It has been found that the profile dimesion at $z=0$ plane is valued to be integer as expected for diffraction patterns. Moving towards the S plane the dimension started to increase and take a fraction values reaching a maxmuim fraction value at the S plane. In contrast to Ref. \cite{Loaiza-et-al-2003,Loaiza-2005} claimed that the fractal structured occurs at the apretured mirror, our results confirm that the fractals occured at the imaging S plane.

Finally, we investigated the generation of these fractal modes in several laser resonator configurations of differing magnification factors (and Fresnel numbers), as detailed in Table~\ref{parameters}, with the results shown in Fig.~\ref{collage}. % \Note{We find that xxxxx results on the fractal dimension xxx.}
Most of the lasers had hexagonal apertures, but we also experimented with an aperture in the shape of the 3rd iteration of the Koch-snowflake, % (Fig.\ \ref{f:KochAperture}), 
a shape approximating a fractal.
We do not analyse the intensity cross-sections in the laser with the Koch-snowflake aperture in any detail here, but note that it is not surprising that the resulting diffraction patterns are fractals \cite{Berry-1979} as there are \emph{several} mechanisms\footnote{Here there are two mechanisms, specifically the MOM effect and diffraction behind a fractal aperture.  This could easily be extended to \emph{three} by periodically repeating the fractal aperture shape across the aperture plane, thereby adding the fractal Talbot effect \cite{Berry-Klein-1996a} to the mix.} at work that all simultaneously shape the eigenmode intensity cross-sections into fractals.
% This investigation is beyond the scope of this paper, so we simply present a few intensity cross-sections here.

%The effect of the magnification of the laser resonator on the dimension of the fractal has been studied. In Fig[XXX], we measured the dimension of two different fractal beams generated by the same apreture from two resonators of magnifications $M=2$ and M=3.33. [Hend add the result ] 

\begin{table}
	%\begin{tabular}{|c|c| c| c| c| }\hline
	\begin{center}
              \begin{tabular*}{0.48\textwidth}{|c| @{\extracolsep{\fill}} c|c|c|c|} \hline
		$R_1$ & 200 mm&  200 mm   & 150 mm & 100 mm\\  \hline
		$R_2$&400 mm&  500 mm & 500 mm &500 mm \\ \hline
		% $g_1g_2$& 1.523& 1.523&   1.65 &1.9 \\ \hline
                      $M$& 2.0 & 2.5 & 3.3 & 5.0 \\ \hline
	\end{tabular*}
	\end{center}
	\caption{Resonator parameters used to design fractal cavities of various magnifications. % and Fresnel numbers.
	}
	\label{parameters}
\end{table}

%%%%%
%%%%%
\section{Discussion and Conclusion}
%%%%%
%%%%%

When not considered in the context of resonators, the existence of 3D self-similar fractal light fields is surprising: the 3D intensity distribution of any light field is fully determined by any transverse cross-section, and so the lowest-loss eigenmode is fully determined by its cross-section in the magnified self-conjugate plane. The existence of self-similar transverse cross-sections whose corresponding beams --- their 3D diffraction patterns --- are also self-similar is far from obvious.  While we have attempted to observe this experimentally, the experimental requirements on imaging are at present prohibitive.  In the 2D case we have been able to confirm the emergence of fractal light from carefully constructed lasers. We have shown experimentally that fractals can be created directly from such laser cavities, confirming a theoretical prediction of some decades.  While the experimental confirmation of 2D fractals reported here concludes an open question in the community, the extension of the theory to 3D opens new exciting avenues for further exploration.

\appendix

\section{Resonator Simulations}
\label{a:simulations}
The simulations of the resonator, shown in Figs \ref{transverse-intensity-distribution-3D-resonator-figure} to \ref{f:selfSimilarity}, were performed using the open-source~\cite{YoungTIM-source} package \emph{Young TIM}, available at \cite{Courtial-et-al-2018c} as a runnable Java Archive, together with a very brief user guide.

Young TIM represents a transverse cross-section through a monochromatic light beam on a rectangular regular grid of points covering a rectangular area in a transverse plane.
At each point, the complex electric field is represented by a complex number $u$, enabling representation of a the amplitude, which is $|u|$; the phase, which is $\mathrm{arg}(u)$; and the intensity, which is $|u|^2$.
The software assumes uniform polarization across the beam, which is a good approximation in the paraxial limit, in which we operate here.
The complexity of the simulation is limited by the memory requirements of calculating a new transverse beam cross-section while storing others.

Propagation from one transverse plane into another through empty space is performed using a standard Fourier algorithm \cite{Sziklas-Siegman-1975} (but without using the Fresnel approximation to simplify the expression for the $z$ component of the wave vector).
This algorithm performs a plane-wave decomposition of the beam, and calculates the sum of all individual plane-wave components in the new transverse plane.
Transmission through optical elements such as apertures and lenses is simulated by multiplying the complex numbers that represent the beam by a position-dependent factor that represents a change in amplitude (in the case of apertures), a change in phase (in the case of lenses), or both.

Repeated propagation through a laser resonator is performed by treating each mirror like a lens of the same focal length, and --- despite the fact that the propagation direction is reversed by reflection off a mirror --- propagating the beam always through a positive distance.

Young TIM has a number of special additional features; perhaps most relevant to this work is the ability to calculate measures of the self-similarity (see App.\ \ref{a:nsed}) of the intensity-cross-section of the beam in the laser resonator.

\section{Self-similarity and Normalised Squared Euclidean Distance}
\label{a:nsed}
In Figs \ref{f:selfSimilarity} and \ref{f:selfSimilarityExperiment} we plot the negative normalised squared Euclidean distance, $-d^2$, between the centres of unstretched and stretched intensity distributions.
We interpret this as a measure of the self-similarity of the centre of intensity cross-sections.

The normalised squared Euclidean distance, $d^2$, is a measure of the difference between two intensity distributions.
If the two intensity distributions are $I_1(x_i,y_j)$ and $I_2(x_i,y_j)$, $d^2$ is defined as
\begin{equation}
d^2 = \frac{1}{2} \frac{\left| (I_1 - \overline{I_1}) - (I_2 - \overline{I_2}) \right|^2}{\left| I_1 - \overline{I_1} \right|^2 + \left| I_2 - \overline{I_2} \right|^2},
\end{equation}
where $| I |^2 = \sum_{i,j} I(x_i,y_j)^2$, $\overline{I} = (\sum_{i,j} I(x_i,y_j))/(N_x N_y)$, and where all sums are over $N_x$ by $N_y$ values of $x_i$ and $y_j$.

In all cases, the unstretched intensity distribution, $I_1$, was known on a discrete grid of points, and it was required to calculate the stretched (by a factor $M$, the transverse magnification) intensity distribution, $I_2$, on the same grid.
For a particular grid position $(x_i, y_j)$ we calculated the value of the stretched intensity distribution there as
\begin{equation}
I_2(x_i,y_j) = I_1(x_i/M, y_j/M),
\end{equation}
whereby the stretching is relative to the origin.
In general, the position $(x_i/M, y_j/M)$ does not coincide with one of the grid positions on which the unstretched intensity distribution was known;
we used bilinear interpolation between the intensity at the four neighbouring represented positions
\cite{Press-et-al-1992-interpolation-in-two-or-more-dimensions} to approximate this value.

It is clear that stretching the unstretched and stretched intensity distributions are not similar far away from the centre, where the unstretched intensity is close to zero (compare, for example, the frames marked ``$\times 1$'' and ``$\times 2$'' in Fig.\ \ref{transverse-intensity-distribution-3D-resonator-figure}).
For this reason, we sum only over the centre of each pattern, specifically the central rectangle of width and height $1/4$ of the width and/or height of the area on which the unstretched intensity cross-section is represented.

\section{Experimental data}
\label{a:experimentalData}
Raw data representing the experimentally obtained intensity on a square grid of points in different transverse planes in a number of resonators are available at~\cite{Courtial-et-al-2018c}.
Two \emph{Mathematica} documents, also available at \cite{Courtial-et-al-2018c}, were used to visualise and evaluate these data:
\begin{enumerate}
\item \texttt{plot log(intensity) images.nb} was used to plot the individual images shown in Fig.~\ref{collage}.
\item \texttt{analysis.nb} was used to calculate the images shown in Fig.\ \ref{f:selfSimilarityZ23To26} and the curve showing the evolution of the self-similarity upon propagation of experimental intensity cross-sections (Fig.\ \ref{f:selfSimilarityExperiment}).
\end{enumerate}

\section{Calculation of the Parameters of the Magnified Self-conjugate Plane}
\label{a:S-plane}

Here we calculate the $Z$ coordinate of the magnified self-conjugate plane S, $Z_S$, and the transverse round-trip magnification, $M$, for the non-confocal cavities used in our experiment.

Our cavities contain a gain medium with refractive index $n$ and geometrical length $g$.
When seen from a paraxial direction, this gain medium \emph{appears} to be of length $g/n$, so it appears to be a distance $g - g/n$ shorter than it actually is.
The cavity itself therefore appears to be shorter by the same distance, so the presence of the gain medium has the effect that the cavity has an effective length
\begin{equation}
L = G - \left( g - \frac{g}{n} \right) = G - \left( 1 - \frac{1}{n} \right) g.
\end{equation}
From now on, all lengths considered are effective (that is, apparent) lengths.

To calculate the position of the plane S, we consider successive imaging of the self-conjugate plane S due to mirror M$_1$ and due to mirror M$_2$.
Imaging of S by M$_1$ into the intermediate image plane $Z = Z_i$ follows the equation
\begin{equation}
\frac{1}{L-Z_S} + \frac{1}{L-Z_i} = \frac{1}{f},
\label{e:M1-imaging}
\end{equation}
where $L-Z_S$ and $L-Z_i$ are the object and image distances, respectively; the transverse magnification is
\begin{equation}
M_{T,1} = - \frac{L-Z_i}{L-Z_S}.
\label{e:M1-magnification}
\end{equation}
Similarly, M$_2$ images the intermediate image plane to the final image plane $Z = Z_f$ according to the equation
\begin{equation}
\frac{1}{Z_i} + \frac{1}{Z_f} = \frac{1}{F}
\label{e:M2-imaging}
\end{equation}
with transverse magnification
\begin{equation}
M_{T,2} = - \frac{Z_f}{Z_i}.
\label{e:M2-magnification}
\end{equation}
The overall transverse magnification of the final image is then
\begin{equation}
M = M_{T,1} M_{T,2}.
\end{equation}
As it is conjugate to itself, the image of S must be in the plane S again, and so
\begin{equation}
Z_f = Z_S.
\label{e:self-conjugate}
\end{equation}

It is straightforward to show that, in the confocal case ($L = F + f$),
\begin{equation}
Z_S = F, \quad
M = - \frac{F}{f}.
\end{equation}
In the non-confocal case,
\begin{equation}
Z_S = \frac{2 f L - L^2 + \sqrt{d}}{2 (f + F - L)}.
\label{e:Z_S}
\end{equation}
and
\begin{equation}
M = \frac{2 f F}{2 f F - 2 f L - 2 F L + L^2 + \sqrt{d}},
\label{e:M}
\end{equation}
where
\begin{equation}
d = L (L - 2 f) (L - 2 F) (L - 2 f - 2 F).
\end{equation}

%\subsection{Evolution of the Intensity Distribution over the Length of the Resonator}
%\label{a:evolution}
%
%\begin{figure*}[ht]
%	\begin{center} \includegraphics[width=\linewidth]{fracz.pdf} \end{center}
%	\caption{Evolution of the eigenmode intensity cross-section between mirrors M$_2$ ($Z = 0\,\mathrm{mm}$) and M$_1$ ($Z = 325\,\mathrm{mm}$).  The magnification was $M$ = 3.32 ($F=250$ \,mm, $f = 75$\,mm) with an aperture of circumradius $r_0 = 2\,$mm.  The geometrical length of the cavity was $G = 325 $\,mm, corresponding to an effective length of $L = 308.5$\,mm due to the refractive index of the gain medium (see Appendix~\ref{a:S-plane}).}
%	\label{fracz}
%\end{figure*}
%
%Fig.\ \ref{fracz} shows measured intensity cross-sections in a number of transverse planes in the resonator.
%The self-conjugate plane S corresponds to $Z \approx 231.8$\,mm, the mirrors were located at $Z = 0$ and $Z = 325$\:mm.

\end{document}